\documentclass[sigconf,natbib=true,anonymous=false]{acmart}
\usepackage{graphicx}
\usepackage{subcaption}
\usepackage{float}
\AtBeginDocument{%
  }
\setcopyright{acmlicensed}
\copyrightyear{2025}
\acmYear{2025}
\acmDOI{XXXXXXX.XXXXXXX}
\acmConference[RecSys EARL '25]{}{September 22--26,
  2025}{Prague, Czech Republic}
\acmISBN{978-1-4503-XXXX-X/2018/06}

\begin{document}
\title{ELIXIR: Efficient and LIghtweight model for eXplaIning  Recommendations}
\author{Ben Kabongo}
%\authornote{Work done during an internship at AgroParisTech and Onepoint.}
\affiliation{%
  \institution{Sorbonne University, CNRS, ISIR}
  \city{Paris}
  \country{France}
}
\email{ben.kabongo@sorbonne-universite.fr}

\author{Vincent Guigue}
\affiliation{%
  \institution{AgroParisTech, UMR MIA Paris-Saclay}
  \city{Palaiseau}
  \country{France}
}
\email{vincent.guigue@agroparistech.fr}

\author{Pirmin Lemberger}
\affiliation{%
  \institution{Onepoint}
    \city{Paris}
  \country{France}
}
\email{p.lemberger@groupeonepoint.com}

\renewcommand{\shortauthors}{Ben Kabongo, Vincent Guigue, Pirmin Lemberger}

\begin{abstract}
Collaborative filtering drives many successful recommender systems but struggles with fine-grained user-item interactions and explainability. As users increasingly seek transparent recommendations, generating textual explanations through language models has become a critical research area.
Existing methods employ either RNNs or Transformers. However, RNN-based approaches fail to leverage the capabilities of pre-trained Transformer models, whereas Transformer-based methods often suffer from suboptimal adaptation and neglect aspect modeling, which is crucial for personalized explanations.
We propose ELIXIR (\textbf{E}fficient and \textbf{LI}ghtweight model for e\textbf{X}pla\textbf{I}ning  \textbf{R}ecommendations), a multi-task model combining rating prediction with personalized review generation. ELIXIR jointly learns global and aspect-specific representations of users and items, optimizing overall rating, aspect-level ratings, and review generation, with personalized attention to emphasize aspect importance.
Based on a T5-small (60M) model, we demonstrate the effectiveness of our aspect-based architecture in guiding text generation in a personalized context, where state-of-the-art approaches exploit much larger models but fail to match user preferences as well.
Experimental results on TripAdvisor and RateBeer demonstrate that ELIXIR significantly outperforms strong baseline models, especially in review generation.
\end{abstract}


\newcommand\vg[1]{\textcolor{black}{#1}}
\newcommand\bk[1]{\textcolor{black}{#1}}

\maketitle

\section{INTRODUCTION}
\begin{figure}[h]
    \centering
    \vspace*{-5mm}
    \includegraphics[width=1\linewidth]{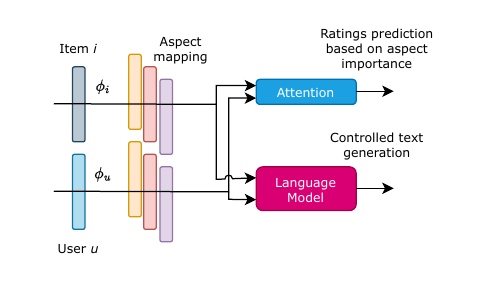}
    \vspace*{-10mm}
    \caption{Schematic overview of our approach: starting from user (u) and item (i) inputs, aspect mapping extracts fine-grained features which are then weighted using a personalized attention mechanism. These refined representations drive both accurate ratings prediction—by emphasizing aspect importance—and controlled review generation via a language model.}
    \label{fig:teaser}
\end{figure}
Many successful recommendation systems rely on collaborative filtering (CF), which learns user preferences and item characteristics from interaction data \cite{koren2009collaborative, koren2021advances, papadakis2022collaborative}. Among CF methods, latent factor models based on matrix factorization (MF) \cite{koren2009matrix, he2017neural, mnih2007probabilistic} have shown strong performance. However, these models primarily rely on latent user and item representations, limiting their ability to capture fine-grained interactions and resulting in a lack of interpretability and explainability.

Beyond accuracy, explainability has become a central focus in recommender system research. Users increasingly expect not only relevant recommendations but also transparent justifications for these recommendations \cite{ge2024survey, zhang2020explainable}. This has led to a growing interest in leveraging language models to generate textual explanations \cite{geng2023recommendation, ni2019justifying, li2023personalized}, including tips \cite{li2017neural} and reviews \cite{dong2017learning, xie2023factual}.
The underlying assumption is that user-generated reviews encapsulate both overall sentiment and fine-grained opinions on specific aspects of an item, making them natural candidates for explainable recommendations.
Several studies have explored explanation generation using recurrent neural networks (RNNs) \cite{costa2018automatic, dong2017learning, li2017neural, ni2018personalized} and Transformers \cite{li2021personalized, li2023personalized, cheng2023explainable, shimizu2024disentangling}. 
However, RNN-based approaches are penalized by the lack of pre-training \cite{brown2020language, dubey2024llama, raffel2020exploring}. Meanwhile, existing Transformer-based methods often show limited performance gains due to suboptimal adaptation strategies. 
Furthermore, most of these methods often overlook aspect modeling, which is essential for capturing fine-grained user preferences and guiding text generation to combine personalization, relevance and fighting hallucinations.
Aspect-based recommender systems (ABRS) already make it possible to model users' opinions on specific aspects of an item. Aspects are mainly extracted from reviews \cite{cheng2018aspect, chin2018anr, guan2019attentive, hasan2024based} and processed by aspect-based sentiment analysis (ABSA) \cite{liu2022sentiment, zhang2022survey} to build more detailed profiles \cite{bauman2017aspect, chen2016learning, ganu2009beyond, he2015trirank, zhang2014explicit}. 
More generally, aspect modeling enables us to refine a user's overall assessment of an item while detailing their opinions on specific aspects, which constitutes a first level of explanation. But this modeling also allows us to improve the textual explanations generated by incorporating fine-grained information specific to each aspect \cite{ni2019justifying, sun2021unsupervised}.

In this paper, we introduce \textbf{ELIXIR} (\textbf{E}fficient and \textbf{LI}ghtweight model for e\textbf{X}pla\textbf{I}ning  \textbf{R}ecommendations), a multi-task model, composed of two modules: the rating prediction module and the personalized review generation module.
Our approach learns both global and aspect-based representations for each user and item from interaction data, optimizing three predictive objectives: overall rating prediction, aspect ratings prediction, and user review generation.
We have built an original personalized attention mechanism to estimate the importance of different aspects for each user and each item, in addition to affinities.
The review generation module derives a personalized continuous prompt from user and item representations, which is then fed into a language model to generate the review. 
Our approach is particularly parameter-efficient, starting from a mainly frozen pre-trained model where only the continuous prompt from user and item profiles is optimized.
Detailed aspect profiles allow us to generate a complete, personalized explanation of the item recommendation through prompt tuning, a particularly parameter-efficient refining method \cite{hanparameter}. 
An illustration of our approach is shown in Figure \ref{fig:teaser}.

Experiments conducted on two real-world multi-aspect datasets, TripAdvisor and RateBeer with a T5-small language model (60M)~\cite{raffel2020exploring}, show that ELIXIR outperforms the state of the art yet based on LLMs several orders of magnitude larger. 
Ablation studies and empirical analyses further highlight the contributions of personalized attention and aspect modeling.
Our contributions are as follows: 
\begin{itemize}
    \item We propose a unified architecture for aspect-based user and item profile modeling, for overall and aspect-specific ratings prediction, and review generation.
    \item We introduce a parameter-efficient generative architecture that outperforms the state-of-the-art in personalized review generation.
    Our prompt tuning approach on a small pre-trained language model demonstrates its effectiveness in controlling text generation.
    \item We conduct experiments and analyses on two multi-aspect real-world datasets. Our results demonstrate the superiority of our model in both rating prediction and review generation, as well as its potential for adaptation to datasets without explicit aspect annotations.
\end{itemize}

\section{RELATED WORK}
Recommender systems are among the early applications of representation learning, particularly through collaborative filtering (CF) \cite{koren2009matrix, papadakis2022collaborative}. CF-based models capture user preferences by identifying similarities among users or items derived from interaction data. Various architectures have been proposed, ranging from matrix factorization \cite{koren2009matrix, mnih2007probabilistic} to deep learning methods \cite{he2017neural, rendle2020neural}.
However, these methods face challenges such as the sparsity of the interaction matrix and the cold start problem, where limited interactions hinder recommendations for new users or items. Content-based (CB) recommender models \cite{javed2021review, lops2011content} mitigate this issue by incorporating item characteristics, although they tend to recommend only similar items.
Hybrid methods combine different recommendation techniques to overcome the limitations of individual approaches. Integrating additional information about users or items—such as item attributes \cite{cheng2018aspect, chin2018anr} or reviews \cite{shuai2022review, sun2021unsupervised}—can significantly enhance the performance of CF-based models.

More specifically, aspect-based recommendation systems (ABRS) improve personalization by modeling user preferences and item characteristics at a finer level of granularity through specific attributes or aspects.  
A first category of ABRS methods extracts aspects from reviews in an unsupervised manner. For example, ALFM \cite{cheng2018aspect} uses an aspect-based topic model (ATM) to learn a multivariate distribution of topics from the reviews, while ANR \cite{chin2018anr} learns aspect representations for users and items by employing attention mechanisms to focus on the most relevant parts of the reviews.  
On the other hand, a second category of ABRS methods focuses on opinion analysis \cite{bauman2017aspect, chen2016learning}, taking advantage of aspect-based sentiment analysis (ABSA) techniques \cite{zhang2022survey} to extract opinions on various aspects from reviews.  
The key advantage of these methods lies in their greater interpretability and explicability compared to conventional recommendation approaches.

Explainability has become a central focus in research on recommender systems, with growing interest in the exploitation of language models to generate textual explanations \cite{li2017neural, dong2017learning, li2021personalized, li2023personalized, xie2023factual}.  
The first methods were based on recurrent neural networks (RNNs), Att2Seq \cite{dong2017learning} and NRT \cite{li2017neural} being notable examples. The latter, in particular, is a multi-task model that generates explanations while simultaneously predicting the overall rating. Other RNN-based methods integrate aspect modeling to better guide explanation generation \cite{ni2018personalized, ni2019justifying, sun2021unsupervised}.  
Subsequently, several explanation generation models took advantage of Transformers. One of the first Transformers-based multi-task models is PETER \cite{li2021personalized}, which integrates heterogeneous data, including word, user, and item embeddings, for overall rating prediction and explanation generation.  
PEPLER \cite{li2023personalized} is derived from PETER and also uses a pre-trained Transformer, GPT-2 \cite{radford2019language}. This architecture has been extended \cite{cheng2023explainable, raczynski2023problem, shimizu2024disentangling}.  
Despite these advances, performance gains over RNN-based methods remain modest, particularly for generating long explanations. This reflects the limitations of previous Transformers adaptations for generating personalized explanations, as most of these methods neglect aspect modeling. Integrating aspect-specific information improves the quality of the generated explanations \cite{ni2019justifying, sun2021unsupervised}.  
To address this challenge, we propose an approach that combines prompt tuning \cite{lester2021power} with the integration of aspect-aware information to better guide the generation process and improve the quality of personalized explanations.

\section{ELIXIR}
\begin{figure*}[h]
  \centering
  \includegraphics[width=1\textwidth]{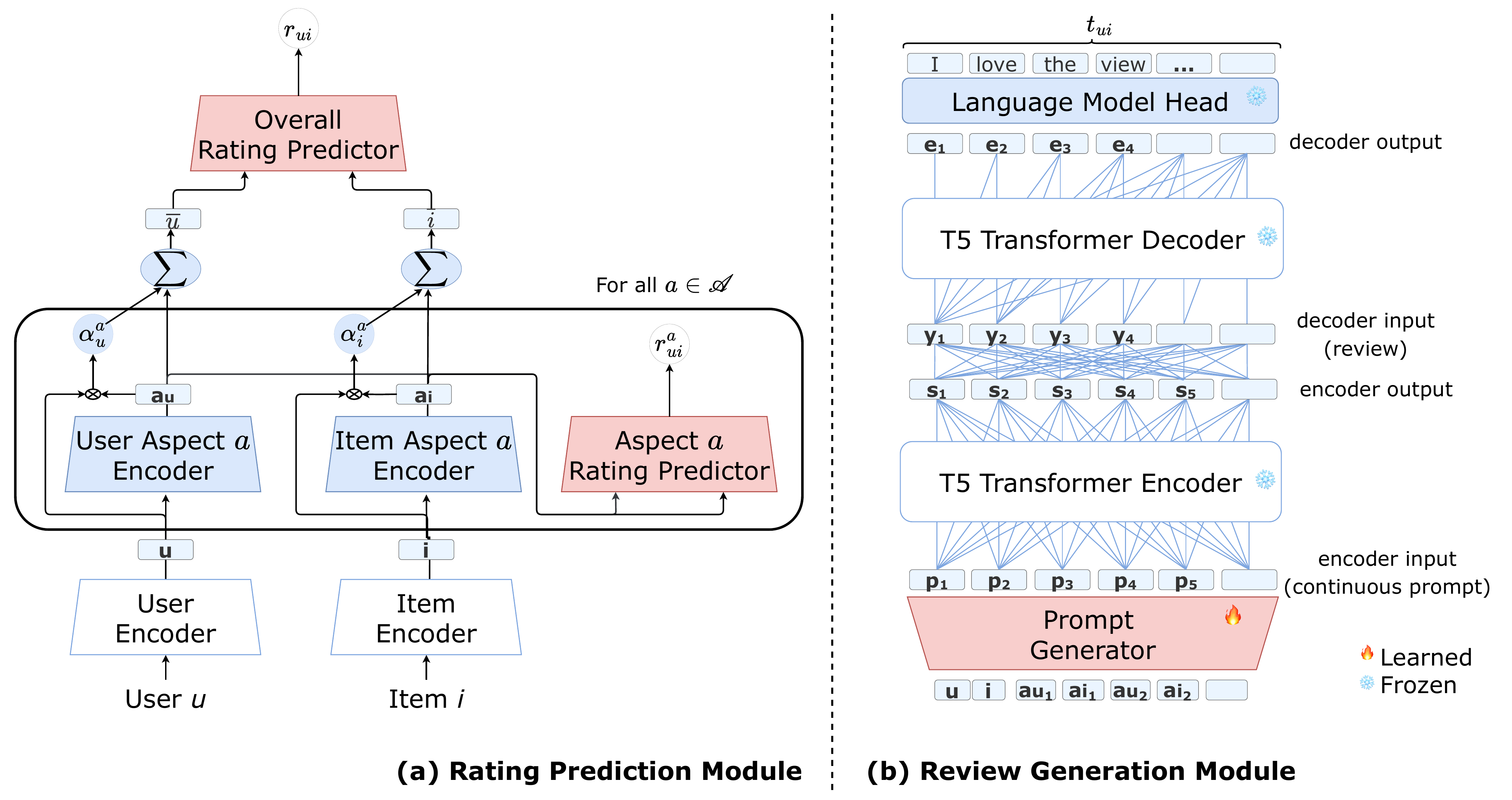}
  \caption{ELIXIR consists of two modules: (a) the rating prediction module on the left and (b) the personalized review generation module on the right. The model learns global representations for both the user and the item, and derives representations for each aspect \(a \in \mathcal{A}\). The rating for aspect \(a\), denoted \(r^a_{ui}\), is obtained from the corresponding user and item aspect representations. The aspect representations are aggregated via personalized attention for both the user and the item to yield the overall rating \(r_{ui}\). Finally, the complete set of global and aspect representations is used to generate a continuous prompt, which is fed to a pre-trained language model (T5) to produce the personalized review \(t_{ui}\).}
  \label{fig:ELIXIR}
\end{figure*}

We propose \textbf{ELIXIR} (\textbf{E}fficient and \textbf{LI}ghtweight model for e\textbf{X}pla\textbf{I}ning  \textbf{R}ecommendations), a multi-task model designed to predict the overall rating, aspect ratings, and personalized review for a given user-item pair.  
ELIXIR consists of two core components: the \textbf{rating prediction module} and the \textbf{personalized review generation module}.  
Figure \ref{fig:ELIXIR} provides an overview of the complete ELIXIR architecture. 

\subsection{Notations and Problem Formulation}
Let $\mathcal{U}$ be the set of users and $\mathcal{I}$ the set of items. 
We denote $|\mathcal{U}|$ and $|\mathcal{I}|$ the number of users and items, respectively.
Let $\mathcal{R}$ be the set of interactions between users and items. 
For an interaction between a user $u$ and an item $i$, let $r_{ui}$ denote the overall rating and $t_{ui}$ the review provided by user $u$ for item $i$.  
The review $t_{ui}$ is represented as a sequence of tokens, $t_{ui} = (y_1, y_2, \dots, y_{|t_{ui}|})$, where each token $y_k$ belongs to the vocabulary set $\mathcal{V}$.  
Given an application domain, let $\mathcal{A}$ denote the set of aspects of interest, and let $|\mathcal{A}|$ represent the number of aspects.  
Aspect ratings for an item $i$ provided by a user $u$ can either be explicitly available or extracted from the review $t_{ui}$ using aspect-based sentiment analysis (ABSA) methods \cite{liu2022sentiment, zhang2021aspect, zhang2022survey}.  
We denote $r^a_{ui}$ as the rating assigned by user $u$ to aspect $a$ of item $i$.  
The interaction set is defined by:
\begin{align}
    \mathcal{R} =  \big\{\big(u, i, r_{ui}, t_{ui}, \{r^a_{ui}\}_{a \in \mathcal{A}} \big)\big\}.
\end{align}
In this article, we focus on an aspect-supervised framework to demonstrate the interest of our generated text control methodology. However, we assume that ABSA techniques have progressed sufficiently in recent years to reasonably consider moving to an unsupervised framework.

Our goal is to learn user and item representations from interaction data, capturing both global and aspect-level interactions to better model overall preferences and fine-grained interests.
Our approach represents user preferences and item characteristics at both levels, using personalized attention to assess the importance of each aspect for individual users and items.
We achieve this through three unified tasks: overall rating prediction, aspect-specific rating prediction, and personalized review generation, ensuring both high performance and explained recommendations.

\subsection{Model}
ELIXIR is composed of two key modules: one for rating prediction and the other for text generation. The first module relies on learned representations at two levels (global --$\mathbf{u}$, $\mathbf{i}$-- and aspect --$\{\mathbf{a}_u\}_{a \in \mathcal{A}}$ and $\{\mathbf{a}_i\}_{a \in \mathcal{A}}$--) and an original personalized attention mechanism to link aspects in the rating prediction. These representations serve as a bridge to the generative module: they are used to build a continuous prompt $\mathbf{p}_{ui}$ to control the generation of personalized text.
We denote $\theta$ as the set of model parameters. 

\subsubsection{Global and Aspect Levels Representations}

\paragraph{Global Level} 
We learn global representations, $\mathbf{u} \in \mathbb{R}^d$ for each user $u \in \mathcal{U}$, and $\mathbf{i} \in \mathbb{R}^d$ for each item $i \in \mathcal{I}$, which encode various information, including user preferences, item characteristics, and other relevant contextual factors.
\paragraph{Aspect Level}  
From the global representations, we aim to extract the preferences of user $u$ and the characteristics of item $i$ concerning aspect $a$.  
To achieve this, we define two functions, $\phi^a_\mathcal{U}, \phi^a_{\mathcal{I}}: \mathbb{R}^d \rightarrow \mathbb{R}^d$, which transform the global representations $\mathbf{u}$ and $\mathbf{i}$ into aspect-specific representations, denoted as $\mathbf{a}_u$ and $\mathbf{a}_i$, respectively.  
These aspect-aware representations are obtained as follows:  
\begin{align}
    \mathbf{a}_u = \phi^a_{\mathcal{U}}(\mathbf{u}), \quad
    \mathbf{a}_i = \phi^a_{\mathcal{I}}(\mathbf{i}), \quad
    \mathbf{a}_u, \mathbf{a}_i \in \mathbb{R}^{d}.
\end{align}
Our experiments led us to choose non-linear projections (MLPs) for $\{\phi^a_{\mathcal{U}},  \phi^a_{\mathcal{I}}\}_{a \in \mathcal{A}}$.
We denote $(\theta_{\mathcal{U}}, \theta_{\mathcal{I}}, \theta_{\mathcal{A}})$ as the set of model parameters corresponding to users, items, and aspects, respectively.

\subsubsection{Personalized Attention}
In personalized recommendation systems, the importance of each aspect varies significantly depending on the user's preferences and the item's characteristics \cite{cheng20183ncf, cheng2018aspect, chin2018anr}. We model these variations in aspect importance using personalized attention, which estimates the weight of each aspect for each user and each item, inspired by attention mechanism from \cite{vaswani2017attention}.
Given a user $u$, we estimate the relative importance weights of aspects, denoted by $\{\alpha^a_u\}_{a \in \mathcal{A}}$, using three linear mapping functions, $q_{\mathcal{U}}, k_{\mathcal{U}}, v_{\mathcal{U}} : \mathbb{R}^d \rightarrow \mathbb{R}^d$. Similarly, for an item $i$, we apply three linear mapping functions, $q_{\mathcal{I}}, k_{\mathcal{I}}, v_{\mathcal{I}} : \mathbb{R}^d \rightarrow \mathbb{R}^d$ to compute the aspect attention weights $\{\alpha^a_i\}_{a \in \mathcal{A}}$.
The importance weights of aspect $a$ for user $u$ and item $i$, respectively denoted by $\alpha^a_u$ and $\alpha^a_i$, are computed as follows:
\begin{align}
    \alpha^a_u = \frac{\text{exp}(q_{\mathcal{U}}(\mathbf{u}) \cdot k_{\mathcal{U}}(\mathbf{a}_u))}{Z_u}, \quad
    \alpha^a_i = \frac{\text{exp}(q_{\mathcal{I}}(\mathbf{i}) \cdot k_{\mathcal{I}}(\mathbf{a}_i))}{Z_i},
\end{align}
%
%where $Z_u = \sum_{a^{'}} \text{exp}(q_{\mathcal{U}}(\mathbf{u}) \cdot k_{\mathcal{U}}(\mathbf{a}^{'}_u))$ and $Z_i = \sum_{a^{'}} \text{exp}(q_{\mathcal{I}}(\mathbf{i}) \cdot k_{\mathcal{I}}(\mathbf{a}^{'}_i))$ are normalization terms.
where $Z_u$ and $Z_i$ are normalization terms.
For a user, the aspect attention weights reflect the relative importance the user assigns to each aspect. For an item, these weights represent the average importance attributed to each aspect by users.

We then compute aggregated representations for the user $u$ and the item $i$, denoted as $\tilde{\mathbf{u}}$ and $\tilde{\mathbf{i}}$, respectively. These representations dynamically capture the user's preferences and the item's characteristics across various aspects. They are obtained as follows:
\begin{align}
    \tilde{\mathbf{u}} = \sum_{a \in \mathcal{A}} \alpha^a_u v_{\mathcal{U}}(\mathbf{a}_u), \quad
    \tilde{\mathbf{i}} = \sum_{a \in \mathcal{A}} \alpha^a_i v_{\mathcal{I}}(\mathbf{a}_i), \quad
    \tilde{\mathbf{u}}, \tilde{\mathbf{i}} \in \mathbb{R}^d.
\end{align}

\subsubsection{Rating Prediction Module}  
The objective of the rating prediction module is to estimate both the overall rating $r_{ui}$ and the aspect-specific ratings $\{r^a_{ui}\}_{a \in \mathcal{A}}$ for a user $u$ and an item $i$. 

\paragraph{Overall Rating Prediction}  
To predict the overall rating $r_{ui}$ of user $u$ for item $i$, we define a function $f: \mathbb{R}^{2 \times d} \rightarrow \mathbb{R}$. The inputs to this function are the aggregated representations of the user and the item, denoted $\tilde{\mathbf{u}}$ and $\tilde{\mathbf{i}}$, respectively. These representations encapsulate the essential user preferences and item characteristics while dynamically integrating the relative importance of different aspects. The overall rating is predicted as follows:
\begin{align}
    \hat{r}_{ui} = f(\tilde{\mathbf{u}}, \tilde{\mathbf{i}}).
\end{align}

\paragraph{Aspect Rating Prediction}  
To predict the aspect-specific rating $r^a_{ui}$, which represents user $u$'s evaluation of aspect $a$ for item $i$, we define a function $g_a: \mathbb{R}^{2 \times d} \rightarrow \mathbb{R}$. The inputs to this function are the representations of aspect $a$ for the user and the item, denoted $\mathbf{a}_u$ and $\mathbf{a}_i$, respectively. These representations capture fine-grained user preferences and item characteristics for aspect $a$. The prediction of the aspect rating is given by:
\begin{align}
    \hat{r}^a_{ui} = g_a(\mathbf{a}_u, \mathbf{a}_i).
\end{align}

We denote $\theta_R$ as the set of parameters specific to the rating prediction module. Our experiments have led us to take non-linear functions (MLPs) for $f$ and $\{g_a\}_{a \in \mathcal{A}}$.

\subsubsection{Personalized Review Generation}
The generative module aims to construct a personalized text explaining recommendations, learned from users' past reviews. Our parameter-efficient approach operates in two steps: (1) generate $\mathbf{p}_{ui}$, a set of continuous representations from the aspect profiles (user and item); (2) optimize these token representations (prompt tuning) to match the target text $t_{ui}$ while keeping the language model frozen.
We denote $\theta_P$ as the set of model parameters for prompt generation. 

\paragraph{Personalized Prompt Generation}
From the global representations $\mathbf{u}$ and $\mathbf{i}$ of user $u$ and item~$i$, along with their aspect representations $\{\mathbf{a}_u, \mathbf{a}_i\}_{a \in \mathcal{A}}$, we employ a function $\psi: \mathbb{R}^{2(1+|\mathcal{A}|) \times d} \rightarrow \mathbb{R}^{\eta \times d_w}$ to obtain the continuous personalized prompt. 
Here, $\eta$ is a hyperparameter of the model that depends on the language model $\theta_{LM}$.
Our experiments with T5-Small have shown that 50 tokens effectively guide generation.
Our hypothesis is that, in addition to the global representations, the user preferences and item characteristics encapsulated in the aspect-aware representations contribute to more effective guidance for the review generation process.
The personalized prompt, denoted $\mathbf{p}_{ui} \in \mathbb{R}^{\eta \times d_w}$, consists of $\eta$ tokens in the latent space of the language model $\theta_{LM}$ of dimension $d_w$.
This prompt captures the same information as the global and aspect representations of the user and item, and is computed as follows:
\begin{align}
    \mathbf{p}_{ui} = \psi\big(\mathbf{u}, \mathbf{i}, \{\mathbf{a}_u, \mathbf{a}_i\}_{a \in \mathcal{A}}\big).
\end{align}

\paragraph{Review Generation}
After obtaining the continuous personalized prompt $\mathbf{p}_{ui}$ for user $u$ and item $i$, we employ a language model $\theta_{LM}$ to generate the review $t_{ui}$, conditioned on the prompt.
In our experiments, we chose the T5 pre-trained model for $\theta_{LM}$.
The generation of the review $t_{ui}$ for user $u$ and item $i$, conditioned on the personalized prompt $\mathbf{p}_{ui}$, is formulated as follows:
\begin{align}
    P_{\theta_P, \theta_{LM}}\big(t_{ui} | \mathbf{p}_{ui}\big) &= 
    P_{\theta_P, \theta_{LM}}\big((y_1, y_2, \dots, y_{|t_{ui}|}) | \mathbf{p}_{ui}\big) \notag \\
    &= \prod_{k=1}^{|t_{ui}|} P_{\theta_P, \theta_{LM}}\big(y_k | \mathbf{p}_{ui}, y_{<k}\big).
\end{align}

\subsection{Optimization}
In this section, we discuss the optimization of the model, covering the loss functions and the training procedure.

\subsubsection{Loss Functions}
The set of model parameters, denoted as $\theta = (\theta_{\mathcal{U}}, \theta_{\mathcal{I}}, \theta_{\mathcal{A}}, \theta_R, \theta_P, \theta_{LM})$, is optimized by minimizing three loss functions corresponding to the three objectives: overall rating prediction, aspect ratings prediction, and review generation.

For overall rating prediction, we employ the mean squared error (MSE) loss, defined as:
\begin{align}
    \label{eq:overall_rating_loss}
    \mathcal{L}_R = \frac{1}{|\mathcal{R}|} \sum_{(u, i) \in \mathcal{R}} \big( r_{ui} - \hat{r}_{ui} \big)^2.
\end{align}
For aspect ratings prediction, we use the average of the mean squared error (MSE) losses computed for each aspect:
\begin{align}
    \label{eq:aspects_rating_loss}
    \mathcal{L}_A = \frac{1}{|\mathcal{A}|} \sum_{a \in \mathcal{A}} \frac{1}{|\mathcal{R}|} \sum_{(u, i) \in \mathcal{R}} \big( r^a_{ui} - \hat{r}^a_{ui} \big)^2.
\end{align}
The total loss of the rating prediction module is a weighted combination of these two losses:
\begin{align}
    \label{eq:rating_loss}
    \mathcal{L}_{rating} = \alpha \mathcal{L}_R + (1 - \alpha) \mathcal{L}_A,
\end{align}
where $\alpha$ is an hyperparameter that control the relative importance of the overall rating loss ($\mathcal{L}_R$) and the aspect ratings loss ($\mathcal{L}_A$).

For personalized review generation, we use the negative log-likelihood (NLL) loss, given by:
\begin{align}
    \label{eq:review_loss}
    \mathcal{L}_{review} = - \frac{1}{|\mathcal{R}|} \sum_{(u, i) \in \mathcal{R}} \frac{1}{|t_{ui}|} \sum_{k=1}^{|t_{ui}|} \log P_{\theta_P, \theta_{LM}} \big( y_k \mid \mathbf{p}_{ui}, y_{<k} \big).
\end{align}

\subsubsection{Training}
Our method naturally aligns with prompt tuning \cite{lester2021power}, allowing us to effectively leverage the pre-trained knowledge. In this configuration, we freeze the language model parameters $\theta_{LM}$ during training and only optimize the prompt generation parameters $\theta_P$. This strategy significantly reduces the number of trainable parameters, thus preserving the rich knowledge encoded within the pre-trained model. 
To maximize efficiency in this setting, we adopt a sequential training approach. Since the personalized prompt depends directly on global and aspect-aware user and item representations, we first optimize these representations by training the rating prediction module and minimizing the loss $\mathcal{L}_{rating}$. Once the learned representations are sufficiently informative, we then train only the prompt generation parameters in the personalized review generation module by minimizing the review generation loss $\mathcal{L}_{review}$, keeping the language model parameters frozen. %The sequential approach is particularly advantageous when the review generation module is trained via prompt tuning.

\section{EXPERIMENTS}

\subsection{Implementation}
We implement key functions in ELIXIR\footnote{Our implementation is available here \url{https://github.com/BenKabongo25/aspect_explainable_recommender}.} using multi-layer perceptrons (MLPs). These include $f$ for the overall rating prediction, $\{g_a\}_{a \in \mathcal{A}}$ for the aspect ratings prediction, $\{\phi^a_{\mathcal{U}}, \phi^a_{\mathcal{I}}\}_{a \in \mathcal{A}}$ for the aspect representations, and $\psi$ for the personalised prompt generation.
We conducted preliminary experiments by implementing the functions $f$ and $\{g_a\}_{a \in \mathcal{A}}$ with a scalar product and the functions $\{\phi^a_{\mathcal{U}}, \phi^a_{\mathcal{I}}\}_{a \in \mathcal{A}}$ and $\psi$ with linear projectors. However, the performance of the model is inferior compared to the performance obtained by implementing these functions with MLPs, suggesting that non-linearity is necessary to translate the different inputs into the different semantic spaces.
For personalized user attention ($q_{\mathcal{U}}, k_{\mathcal{U}}, v_{\mathcal{U}}$) and item attention ($q_{\mathcal{I}}, k_{\mathcal{I}}, v_{\mathcal{I}}$) functions, we learn separate projection matrices. For example, for the function $q_{\mathcal{U}}$,  we use a projection matrix $\mathbf{W}^q_{\mathcal{U}} \in \mathbb{R}^{d \times d}$.
The generation of the review is given by $P_{\theta_P, \theta_{LM}}\big(t_{ui} | \mathbf{p}_{ui}\big)$. 
In our implementation of ELIXIR, we utilize the pre-trained Transformer-based T5 language model \cite{raffel2020exploring}, and our experiments follow the prompt tuning approach within a sequential training setup. The hyperparameter $\eta$, which specifies the number of prompt tokens, depends on the language model, as demonstrated in \cite{lester2021power}.
%.

\subsection{Ablations}  
We introduce various ablations of the ELIXIR model to assess the contribution of individual components, such as personalized attention and aspect modeling.

\subsubsection{Personalized Attention}  
We denote \textit{ELIXIR -Attention} as the ablation of the ELIXIR model in which personalized attention is replaced with max pooling.
The aggregated aspect representations of user $u$ and item $i$ are computed using a max pooling operation over all their aspect representations, as follows:
\begin{align}
    (\tilde{\mathbf{u}})_j = \max_{a \in \mathcal{A}} (\mathbf{a}_u)_j, \quad (\tilde{\mathbf{i}})_j = \max_{a \in \mathcal{A}} (\mathbf{a}_i)_j, \quad \forall j \in \{1, \dots, d\}.
\end{align}

\subsubsection{Aspect Modeling}  
We denote \textit{ELIXIR -Aspects} as the ablation of the ELIXIR model in which aspect modeling is removed. In this variant, the model does not learn aspect-aware representations for users and items, and we also omit aspect ratings prediction and personalized attention.   
For overall rating prediction, we redefine the function $f: \mathbb{R}^{2 \times d} \rightarrow \mathbb{R}$, and for personalized prompt generation, we redefine the function $\psi: \mathbb{R}^{2 \times d} \rightarrow \mathbb{R}^{\eta \times d_w}$. Since aspect modeling is omitted, these functions take as input only the global representations of the user and item, as follows:
\begin{align}
    \hat{r}_{ui} = f(\mathbf{u}, \mathbf{i}), \quad \mathbf{p}_{ui} = \psi(\mathbf{u}, \mathbf{i}).
\end{align}  

\subsubsection{Global Representations}  
We denote \textit{ELIXIR -Global} as an ablation of the ELIXIR model where overall rating prediction, aspect ratings prediction, and personalized continuous prompt generation rely solely on the global representations of users and items.  
This ablation allows us to assess the combined importance of aspect modeling and personalized attention across all model tasks.  
In this model ablation, the functions $f: \mathbb{R}^{2 \times d} \rightarrow \mathbb{R}$, $g_a: \mathbb{R}^{2 \times d} \rightarrow \mathbb{R}$, and $\psi: \mathbb{R}^{2 \times d} \rightarrow \mathbb{R}^{\eta \times d_w}$ for overall rating prediction, aspect $a$ rating prediction, and prompt generation, are defined as follows:
\begin{align}
    \hat{r}_{ui} = f(\mathbf{u}, \mathbf{i}), \quad \hat{r}^a_{ui} = g_a(\mathbf{u}, \mathbf{i}), \quad \mathbf{p}_{ui} = \psi(\mathbf{u}, \mathbf{i}).
\end{align}

\subsection{Experimental Setup}

\subsubsection{Datasets}
Our experiments are conducted on two real-word multi-aspect datasets:
\begin{itemize} 
    \item \textbf{TripAdvisor} \footnote{\url{https://www.cs.virginia.edu/~hw5x/Data/LARA/TripAdvisor/}} \cite{wang2010latent}: includes reviews covering six aspects of hotels: cleanliness, location, room, service, sleep quality, and value. 
    \item \textbf{RateBeer} \footnote{\url{https://cseweb.ucsd.edu/~jmcauley/datasets.html\#multi_aspect}} \cite{mcauley2012learning}: includes reviews focusing on four aspects of beers: appearance, aroma, palate, and taste. 
\end{itemize}
\begin{table}[h]
      \caption{Datasets description. Statistics after filtering are shown in brackets.}
      \label{tab:dataset_description}
      \begin{tabular}{lrr}
        \toprule
        \textbf{Datasets} & \textbf{TripAdvisor} & \textbf{RateBeer} \\
        \midrule
        Aspects & 6 & 4 \\
        Users & 716 870 (8 830) & 40 213 (8 384) \\
        Items & 10 008 (2 903) & 110 419 (5 093) \\
        Interactions & 1.4M (62.7K) & 2.8M (201.7K) \\
        \bottomrule
      \end{tabular}
\end{table}
For each dataset, we exclude interactions with missing information and retain only users and items that have at least 5 reviews. Table \ref{tab:dataset_description} provides a summary of the datasets before and after applying these filtering steps.
Although the number of datasets is limited, we emphasize the diversity of the topics covered. We are confident in the possibility of rapidly extending these experiments to aspect-unsupervised data using ABSA techniques.

\subsubsection{Evaluation metrics}
We use the Root Mean Squared Error ({RMSE}) and Mean Absolute Error ({MAE}) metrics to evaluate the models on rating prediction.
For review generation, we evaluate the models using text quality metrics, including {METEOR} \cite{banerjee2005meteor}, {BLEU} \cite{papineni2002bleu}, {ROUGE} \cite{lin2004rouge}, and {BERTScore} \cite{zhang2019bertscore}.

\subsubsection{Baselines}
We compare ELIXIR against a diverse set of baselines, including traditional rating prediction methods, aspect-based recommendation models, explanation generation approaches, and multi-task models. To evaluate the contribution of the different components of ELIXIR, we also consider its various ablations, including {ELIXIR -Attention}, {ELIXIR -Aspects}, and {ELIXIR -Global}.

\paragraph{Ratings prediction} We employ the following baselines: {Average}, {MF} \cite{koren2009matrix}, {MLP} \cite{he2017neural}, and {NeuMF} \cite{he2017neural}. The Average method predicts the mean rating across all user-item pairs.
For aspect-based recommendation, we evaluate two baselines: {ALFM} \cite{cheng2018aspect} and {ANR} \cite{chin2018anr}. These methods learn representations of aspects from reviews in an unsupervised manner, making it challenging to align the learned aspects with those in the dataset. We consider the ablations {ELIXIR -Global} and {ELIXIR -Attention} as additional baselines specifically for aspect ratings prediction.

\paragraph{Review generation} We consider two categories of explanation generation models. The first category consists of RNN-based models, including {Att2Seq} \cite{dong2017learning} and the multi-task architecture {NRT} \cite{li2017neural}. The second category includes multi-task Transformer-based models, such as {PETER} \cite{li2021personalized} and {PEPLER} \cite{li2023personalized}. PETER uses an unpretrained Transformer, while PEPLER employs the pre-trained GPT-2 \cite{radford2019language} in a fine-tuning approach.
Most of the text-based explanation generation methods considered rely on latent representations of users and items—typically encoded as just two tokens—to condition the entire explanation.
We exclude models such as those proposed in \cite{cheng2023explainable, raczynski2023problem, shimizu2024disentangling}, which are extensions of PETER and PEPLER, as their performance closely mirrors that of the original models.

\subsubsection{Setup}
Each dataset is split into training, validation, and test sets using an 80:10:10 ratio. All ratings, including overall and aspect ratings, are standardized to a scale of 1 to 5. Review lengths are capped at 128 words for all models. Models are trained on the training set, and hyperparameters are tuned using the validation set. The results reported correspond to the performance of the models on the test set.
For most models, we use the hyperparameters specified in the original papers. Specifically, for {PEPLER} \cite{li2023personalized}, we use GPT-2 (124M) \cite{radford2019language} as detailed in the original paper. For {ELIXIR}, we use the T5-Small model (60M) \cite{raffel2020exploring}, which is half the size of GPT-2.
After preliminary experiments, we set $\eta$, the number of tokens in the personalized prompt, to 50. The model dimension, $d$, is set to 256, while the T5-Small model's dimension is 512 \cite{raffel2020exploring}. The number of layers in the MLPs is fixed at 2, and we use the Rectified Linear Unit (ReLU) \cite{nair2010rectified} as the activation function. The dropout probability is set to 0.1 to prevent overfitting.
We find that setting $\alpha = \frac{1}{|\mathcal{A}| + 1}$, where $|\mathcal{A}|$ represents the number of aspects, strikes the best balance between performance on overall and aspect ratings.
All models are trained for a maximum of 100 epochs. For ELIXIR, we employ sequential training and prompt tuning. 
The rating prediction module is trained for 50 epochs, followed by another 50 epochs for training the review generation module. Training is performed using the Adam \cite{kingma2014adam} optimizer with a learning rate of $10^{-3}$.

\subsection{Ratings prediction}

\subsubsection*{Overall rating}
\begin{table}
      \caption{Performance on overall rating prediction.}
      \label{tab:results_overall_rating_prediction}
      %\resizebox{0.5\textwidth}{!}{
      \begin{tabular}{lcccc}
        \toprule
        & \multicolumn{2}{c}{\textbf{TripAdvisor}} & \multicolumn{2}{c}{\textbf{RateBeer}} \\
        \cmidrule(lr){2-3}  \cmidrule(lr){4-5}
        Model & RMSE $\downarrow$ & MAE $\downarrow$ & RMSE $\downarrow$ & MAE $\downarrow$ \\
        \midrule
        Average & 0.932 & 0.645 & 0.571 & 0.424 \\
        MF & 0.840 & 0.646 & \textbf{{0.411}} & \textbf{{0.300}} \\
        MLP & \underline{0.833} & \underline{0.565} & 0.464 & 0.324 \\
        NeuMF & 0.840 & 0.570 & 0.473 & 0.329 \\
        \midrule
        %A3NCF & 1.0100 & 0.7639 & 0.6085 & 0.4668 \\
        ALFM & 0.896 & 0.691 & 0.433 & 0.314 \\
        ANR & \underline{0.847} & \underline{0.607} & \underline{0.423} & \underline{0.308} \\
        \midrule
        NRT & 0.859 & 0.548 & 0.420 & 0.306 \\
        PETER & 0.807 & 0.532 & \underline{0.415} & \textbf{{0.300}} \\
        PEPLER & \underline{0.779} & \underline{0.478} & 0.430 & 0.305 \\
        \midrule   
        {ELIXIR} & \textbf{{0.748}} & \textbf{{0.447}} & \underline{0.416} & \underline{0.305} \\
        \textit{\quad -Attention} & 0.771 & 0.513 & 0.421 & 0.311 \\
        \textit{\quad -Global} & 0.865 & 0.632 & 0.443 & 0.332 \\
        \bottomrule
      \end{tabular}
      %}
\end{table}
We compare ELIXIR to various baseline models for overall rating prediction, with results reported in Table \ref{tab:results_overall_rating_prediction}. ELIXIR consistently outperforms all baselines across all metrics, particularly on the TripAdvisor dataset. On the RateBeer dataset, it remains among the top-performing models, with performance close to MF and PETER.
The superiority of ELIXIR over traditional and multi-task approaches can be attributed to its integration of aspect modeling. Likewise, its advantage over aspect-based baselines highlights the benefits of supervised modeling, especially through directly extracting opinions on aspects from reviews.
The ablations {ELIXIR -Attention} and {ELIXIR -Global} perform worse than the full version of ELIXIR, confirming the importance of personalized attention and the joint use of global and aspect-aware representations for achieving the highest accuracy.

\subsubsection*{Aspect ratings}
\begin{table*}
      \caption{Performance on aspect ratings prediction. We report the mean and standard deviation of all aspects.}
      \label{tab:results_aspects_ratings_prediction}
      \begin{tabular}{lcccc}
        \toprule
        & \multicolumn{2}{c}{\textbf{TripAdvisor}} & \multicolumn{2}{c}{\textbf{RateBeer}} \\
        \cmidrule(lr){2-3}  \cmidrule(lr){4-5}
        Model & RMSE $\downarrow$ & MAE $\downarrow$ & RMSE $\downarrow$ & MAE $\downarrow$ \\
        \midrule
        Average & 1.014 (0.087) & 0.801 (0.057) & 0.605 (0.011) & 0.489 (0.023) \\
        \midrule
        {ELIXIR} & \textbf{0.753 (0.081)} & \textbf{0.451 (0.053)} & \textbf{0.465 (0.034)} & \textbf{0.354 (0.030)}\\
        \textit{\quad -Attention} & 0.785 (0.073) & 0.554 (0.051) & 0.486 (0.031) & 0.373 (0.030) \\
        \textit{\quad -Global} & 0.860 (0.076) & 0.631 (0.056) & 0.495 (0.035) & 0.383 (0.031) \\
        \bottomrule
      \end{tabular}
\end{table*}
\begin{table*}[h]
      \caption{Performance on review generation (\%).}
      \label{tab:results_review_generation}
      \begin{tabular}{lcccccccc}
        \toprule
        \textbf{TripAdvisor} & METEOR $\uparrow$ & BLEU $\uparrow$ & ROUGE-1 $\uparrow$ & ROUGE-2 $\uparrow$ & ROUGE-L $\uparrow$ & BERT-P $\uparrow$ & BERT-R $\uparrow$ & BERT-F1 $\uparrow$ \\
        \midrule
        Att2Seq & 18.611 & 04.690 & 28.783 & 06.473 & 18.523 & {85.348} & 83.676 & {84.490} \\
        NRT & 17.219 & 03.405 & 25.833 & 05.194 & 17.539 & 82.828 & 81.533 & 82.161 \\
        PETER & 17.955 & 03.943 & 27.974 & 05.906 & 18.252 & 85.037 & 83.823 & 84.406 \\
        PEPLER$_{\text{GPT-2}}$ & {24.340} & {11.400} & {33.831} & {11.679} & {22.452} & 82.635 & {84.945} & 83.726 \\
        \midrule
        {ELIXIR$_{\text{T5-Small}}$} & \textbf{42.752} & \textbf{33.544} & \textbf{53.285} & \textbf{37.878} & \textbf{44.053} & \textbf{90.686} & \textbf{88.478} & \textbf{89.554} \\
        \textit{\quad-Aspects} & 27.642 & 10.029 & 39.076 & 21.970 & 29.594 & 88.001 & 85.193 & 86.540 \\
        \midrule
        \textbf{RateBeer} & METEOR $\uparrow$ & BLEU $\uparrow$ & ROUGE-1 $\uparrow$ & ROUGE-2 $\uparrow$ & ROUGE-L $\uparrow$ & BERT-P $\uparrow$ & BERT-R $\uparrow$ & BERT-F1 $\uparrow$ \\
        \midrule
        Att2Seq & 18.611 & 04.690 & 28.783 & 06.473 & 18.523 & 85.348 & 83.676 & 84.490 \\
        NRT & 24.963 & 08.737 & 32.589 & 11.472 & 26.629 & 85.046 & 82.992 & 83.985 \\
        PETER & {28.818} & {11.518} & {35.504} & {13.620} & {29.668} & {87.340} & 85.621 & 86.448 \\
        PEPLER$_{\text{GPT-2}}$ & 28.266 & 10.143 & 32.444 & 11.182 & 26.248 & 84.020 & 86.063 & 84.990 \\
        \midrule
        {ELIXIR$_{\text{T5-Small}}$} & \textbf{40.763} & \textbf{24.160} & \textbf{46.371} & \textbf{25.818} & \textbf{39.461} & \textbf{90.483} & \textbf{89.135} & \textbf{89.792} \\
        \textit{\quad-Aspects} & 32.675 & 13.652 & 39.068 & 17.106 & 32.464 & 89.365 & 87.323 & 88.310 \\
        \bottomrule
      \end{tabular}
\end{table*}
\begin{table}
      \caption{Impact of the number of tokens in the personalized prompt ($\eta$) on review generation  (TripAdvisor dataset).}
      \label{tab:results_number_tokens}
       \resizebox{0.35\textwidth}{!}{
      \begin{tabular}{lccc}
        \toprule
        $\eta$ & METEOR $\uparrow$ & BLEU $\uparrow$ & ROUGE-2 $\uparrow$ \\
        \midrule
        PEPLER & 24.340 & 11.400 & 11.679 \\ %& 83.7264 \\
        \midrule
        2 & 12.159 & 01.282 & 04.436 \\ %& {84.2862} \\
        5 & 16.730 & 03.546 & 06.046 \\ %& {83.8951} \\
        10 & 21.160 & 07.692 & 10.330 \\ %& {84.6978} \\
        20 & \underline{29.379} & \underline{17.018} & \underline{20.200} \\ %& 86.65443 \\
        50 & \textbf{42.752} & \textbf{33.544} & \textbf{37.878} \\ %& \textbf{89.5549} \\
        \bottomrule
      \end{tabular}
      }
\end{table}
\begin{table*}[h]
    \caption{Visualization of attention on example from TripAdvisor dataset. For each aspect, we show the rating predicted by ELIXIR, along with the actual rating in brackets. The aspect ratings and their relative importance are contrasted with the user's actual review. An alignment is observed between the importance of the aspects and the content of the review.}
    \label{tab:attention_visualisation}
    \centering
    \resizebox{\textwidth}{!}{
    \begin{tabular}{c|c}
        \begin{tabular}{c}
            \includegraphics[width=0.2\textwidth]{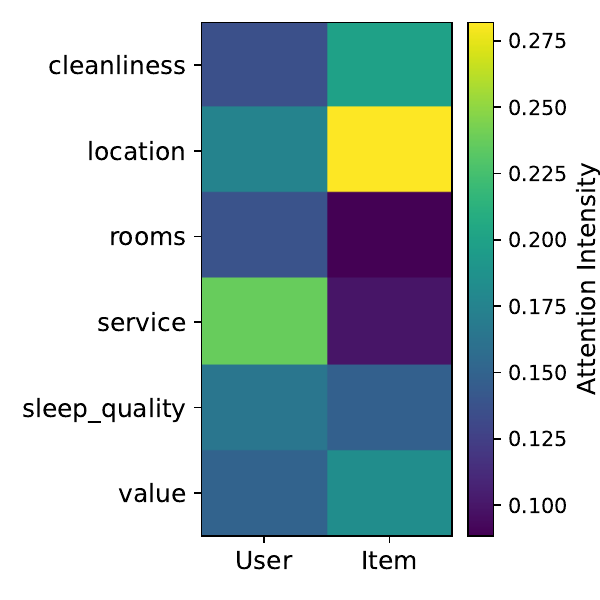}
        \end{tabular} &
        \begin{tabular}{ll|l}
            %\toprule
            \textbf{Aspect} & \textbf{Rating} & \textbf{Ground truth review} \\
            \midrule
            Cleanliness & 4.9 (5.0) & if we go back to paris, we are staying here again. the place is so charming and \\
            Location & 5.0 (5.0) & overlooks the beautiful luxembourg gardens. \underline{the staff were sooo hospitiable.} \\
            Rooms & 5.0 (5.0) &  \underline{always asking what they could do to help us. they arranged two tours for us,} \\
            \underline{Service} & 5.0 (5.0) & \underline{recommended places to eat and then made the reservations for us, arranged} \\
            Sleep & 5.0 (5.0) & \underline{transportation from and to the airport, etc. royce and xavier, i can't thank you} \\
            Value & 5.0 (5.0) & \underline{enough!} also, so many places are in walking distance, like notre dame and the \\
            Overall & 4.9 (5.0) & louvre. you can't help but fall in love with this place! \\
            %\bottomrule
        \end{tabular} \\
    \end{tabular}
    }
\end{table*}
We compare ELIXIR with the Average baseline and the ablations {ELIXIR -Global} and {ELIXIR -Attention} for aspect ratings prediction. For each model, we calculate the RMSE and MAE for all aspects and then report the mean and standard deviation of these metrics across all aspects. The results are presented in Table \ref{tab:results_aspects_ratings_prediction}.
ELIXIR consistently and significantly outperforms all other methods across all metrics and datasets. Both ablations surpass the Average baseline. 
We observe that the {ELIXIR -Global} ablation performs worse than {ELIXIR -Attention}. This performance gap arises because {ELIXIR -Global} relies solely on global representations, whereas {ELIXIR -Attention} retains aspect-specific representations. This clearly demonstrates the value of explicitly learning aspect-aware representations for accurate aspect ratings prediction.
Moreover, the full ELIXIR model consistently outperforms the {ELIXIR -Attention} ablation, which replaces personalized attention with a max pooling operation. This further confirms that personalized attention provides a more effective aggregation strategy for user and item aspect representations than max pooling.
Finally, an additional benefit of ELIXIR over traditional rating prediction or multi-task baselines is its capability to explicitly predict aspect ratings alongside overall rating and personalized review, providing richer and more informative recommendations.

\subsection{Review generation}
We evaluate ELIXIR against various baselines and the {ELIXIR -Aspects} ablation for review generation using text quality metrics, with the results reported in Table \ref{tab:results_review_generation}.
ELIXIR and ELIXIR -Aspects consistently outperform all other models across all metrics and datasets, with ELIXIR achieving superior performance compared to its {ELIXIR -Aspects} ablation. These results highlight the limitations of RNN-based approaches in effectively capturing long-range context, and note that Transformer-based baselines (PETER and PEPLER) do not always significantly outperform RNN-based models, particularly on the TripAdvisor dataset.
The underperformance of the {ELIXIR -Aspects} ablation confirms our hypothesis that incorporating aspect-aware information alongside global representations is crucial for enhancing the quality of generated reviews.

Furthermore, we analyze the impact of $\eta$, the number of tokens in the personalized prompt, on review generation quality for the TripAdvisor dataset. As shown in Table \ref{tab:results_number_tokens}, starting from 20 tokens, ELIXIR outperforms the PEPLER baseline.
The best performance is achieved with 50 tokens.
All baseline methods condition their generation solely on learned user and item representations, typically represented by just two tokens. In contrast, our approach employs a richer, personalized continuous prompt, with its length optimized as a hyperparameter based on the language model used. 
Notably, both ELIXIR -Aspects and the full ELIXIR model outperform the PEPLER model despite using T5-Small, which has roughly half the parameters of the GPT-2 model used by PEPLER. This further underscores the effectiveness and efficiency of our personalized prompt tuning approach.

\subsection{Aspect modeling}

Aspect modeling, by learning aspect-based representations for each user and item, enhances the performance of the model across all tasks.  
For overall and aspect-specific ratings prediction, as reported in Tables \ref{tab:results_overall_rating_prediction} and \ref{tab:results_aspects_ratings_prediction}, both the ELIXIR model and its {ELIXIR -Attention} ablation, which replaces attention with max pooling while preserving aspect-aware representations, significantly outperform the {ELIXIR -Global} ablation, which relies solely on global representations.  
In review generation, the {ELIXIR -Aspects} ablation, which neither learns aspect-aware representations nor predicts aspect ratings, is also considerably outperformed by ELIXIR.  
These results highlight the crucial role of aspect-aware representations in encoding user preferences and item characteristics at the aspect level, leading to more accurate predictions and more informative review generation.

\begin{figure}
    \centering
    \includegraphics[width=0.8\linewidth]{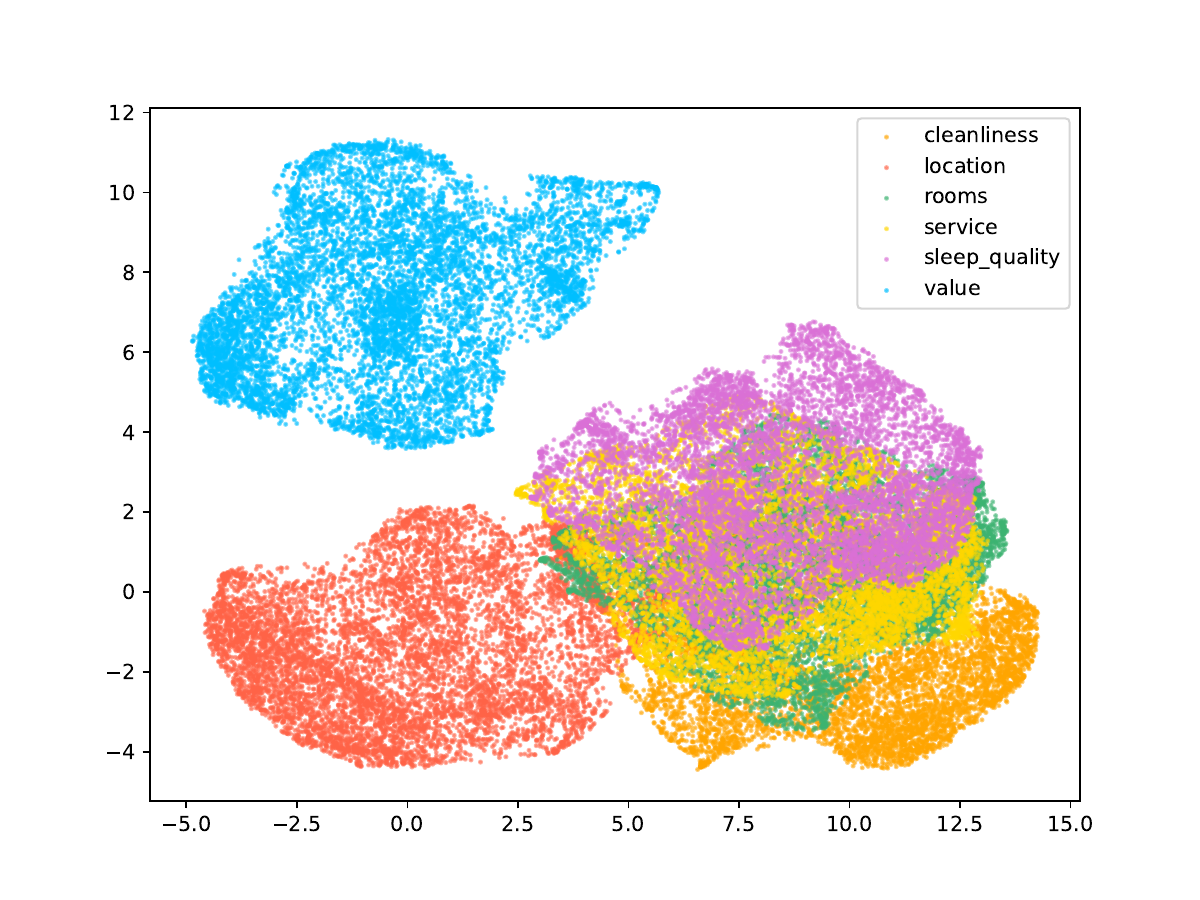}
    \caption{Projection and clustering of user aspect representations for the TripAdvisor dataset.}
    \label{fig:embeddings_plot}
\end{figure}

We conduct an empirical study to examine user representations based on the aspects learned by the model on the TripAdvisor dataset. These representations are projected into a two-dimensional space, and grouped by aspect. The results are presented in Figure \ref{fig:embeddings_plot}.  
The representations based on the aspects learned by ELIXIR allow a coherent separation of users according to the aspects, thus demonstrating that the model effectively captures user preferences according to specific criteria.  
In particular, we observe an overlap between the aspects \textit{rooms, service, sleep quality} and \textit{cleanliness}, which suggests that these aspects are more similar to each other than to \textit{location} and \textit{value}.  
This highlights the advantage of aspect modeling, which allows learning user preferences and item characteristics at a fine-grained level, leading to better structured and more interpretable representations.

\subsection{Personalized Attention}
The personalized attention mechanism in ELIXIR infers the relative importance of aspects for each user and item.
ELIXIR achieves significant performance gains over its ablations ELIXIR -Attention and ELIXIR -Global, particularly in overall rating prediction and aspect rating prediction (Tables \ref{tab:results_overall_rating_prediction} and \ref{tab:results_aspects_ratings_prediction}). Notably, ELIXIR -Attention replaces personalized attention with max pooling, resulting in a performance drop, highlighting that personalized attention better aggregates aspect information than max pooling.
The learned attention weights can be interpreted as proxies for the importance of each aspect to a given user or item. To validate this hypothesis, we analyzed attention weights for user-item pairs and examined their consistency with the actual review content. An example from TripAdvisor dataset is provided in Table \ref{tab:attention_visualisation}.
In this example, the most important aspect for the user, according to attention, is \textit{service}, while for the item, it is \textit{location}. Comparing these attention weights with the review content reveals a strong alignment with the user’s preferences and the aspect importance inferred by ELIXIR. Notably, \textit{service} is the most frequently mentioned aspect in the review, with a positive sentiment, while \textit{location}, the most important aspect for the item, is also emphasized.
These empirical analyses confirm that ELIXIR's personalized attention effectively captures the relative importance of aspects for users and items. Beyond predicting overall and aspect ratings and generating reviews, the inferred aspect importance can serve as an additional explanatory factor, enhancing the interpretability of recommendations.

\section{LIMITATIONS AND FUTURE WORKS}
ELIXIR shows promising performance, but some limitations remain. Currently, the model relies on aspect annotations, which restricts its applicability in contexts where such annotations are unavailable or vary in quality. 
The experimental evaluation, limited to two specific domains, as well as the use of a small-scale model (T5-Small), may also constrain the generalizability and the quality of the generated explanations.
To address these issues, future work will aim to develop unsupervised ABSA methods that automatically extract aspects from unannotated datasets, thereby making the approach more adaptable. Moreover, exploring hybrid architectures or employing larger language models that combine full fine-tuning with prompt tuning could improve the quality and fluency of the generated reviews. Integrating visual data and metadata represents another important avenue to enrich the system’s contextual explainability. Finally, incorporating user feedback mechanisms could help refine personalization and allow the recommendations and their explanations to adapt in real time to users’ evolving preferences.

\section{CONCLUSION}
In this paper, we introduce ELIXIR, a multi-task model composed of a rating prediction module —for overall and aspect ratings— and a personalized review generation module. Our approach jointly learns global and aspect-based representations for users and items, and derives a personalized continuous prompt in the semantic space of a language model. Personalized attention enables us to infer the relative importance of aspects for users and items, supporting our hypothesis that integrating aspect information enhances both recommendations and their explanations.
We implement ELIXIR using the pre-trained T5 language model, employing prompt tuning and sequential training to leverage pre-training effectively. Our experiments on two real-world multi-aspect datasets, TripAdvisor and RateBeer, demonstrate the superiority of ELIXIR over strong baselines across all tasks, particularly in review generation. Empirical analyses and ablation studies confirm the significant contributions of each model component, especially aspect modeling and personalized attention.

\bibliographystyle{ACM-Reference-Format}
\bibliography{references}
\end{document}